\documentclass[aps,prl,amsmath,amssymb,reprint,superscriptaddress,nofootinbib]{revtex4-2}
\usepackage{dutchcal}
\usepackage{graphicx}
\usepackage{dcolumn}
\AtBeginDocument{\usepackage{booktabs}}
\usepackage{units}
\usepackage{soul}
\usepackage{braket}
\usepackage{placeins}
\usepackage[breaklinks=true,colorlinks,citecolor=blue,linkcolor=blue,urlcolor=blue]{hyperref}
\usepackage[]{lineno}
\usepackage{color}
\definecolor{red}{RGB}{255,0,0}

\usepackage{comment}
\usepackage{enumitem}
\renewcommand{\ket}[1]{\vert #1 \rangle}
\renewcommand{\bra}[1]{\langle #1 \vert}

\begin{document}
	

\newcommand{\thetitle}{A high-performance quantum memory for quantum interconnects}

\author{Hao-Xuan Luo}
\thanks{These authors contributed equally to this work.}
\affiliation {Key Laboratory of Atomic and Subatomic Structure and Quantum Control (Ministry of Education), Guangdong Basic Research Center of Excellence for Structure and Fundamental Interactions of Matter, School of Physics, South China Normal University, Guangzhou 510006, China} 

\affiliation {Guangdong Provincial Key Laboratory of Quantum Engineering and Quantum Materials, Guangdong-Hong Kong Joint Laboratory of Quantum Matter, South China Normal University, Guangzhou 510006, China}

\author{Chang Li}
\email[]{lichangphy@gmail.com}
\affiliation {Key Laboratory of Atomic and Subatomic Structure and Quantum Control (Ministry of Education), Guangdong Basic Research Center of Excellence for Structure and Fundamental Interactions of Matter, School of Physics, South China Normal University, Guangzhou 510006, China} 

\affiliation {Guangdong Provincial Key Laboratory of Quantum Engineering and Quantum Materials, Guangdong-Hong Kong Joint Laboratory of Quantum Matter, South China Normal University, Guangzhou 510006, China}
 
\affiliation{Quantum Science Center of Guangdong-Hong Kong-Macao Greater Bay Area, Shenzhen, China}

\author{$^{*\,}$ Jia-Ling Ren}
\thanks{These authors contributed equally to this work.}
\affiliation {Key Laboratory of Atomic and Subatomic Structure and Quantum Control (Ministry of Education), Guangdong Basic Research Center of Excellence for Structure and Fundamental Interactions of Matter, School of Physics, South China Normal University, Guangzhou 510006, China} 

\affiliation {Guangdong Provincial Key Laboratory of Quantum Engineering and Quantum Materials, Guangdong-Hong Kong Joint Laboratory of Quantum Matter, South China Normal University, Guangzhou 510006, China}

\author{Yuan Yuan} 
\affiliation {Key Laboratory of Atomic and Subatomic Structure and Quantum Control (Ministry of Education), Guangdong Basic Research Center of Excellence for Structure and Fundamental Interactions of Matter, School of Physics, South China Normal University, Guangzhou 510006, China} 

\affiliation {Guangdong Provincial Key Laboratory of Quantum Engineering and Quantum Materials, Guangdong-Hong Kong Joint Laboratory of Quantum Matter, South China Normal University, Guangzhou 510006, China}

\author{Yong-Li Wen}
\affiliation {Key Laboratory of Atomic and Subatomic Structure and Quantum Control (Ministry of Education), Guangdong Basic Research Center of Excellence for Structure and Fundamental Interactions of Matter, School of Physics, South China Normal University, Guangzhou 510006, China} 

\affiliation {Guangdong Provincial Key Laboratory of Quantum Engineering and Quantum Materials, Guangdong-Hong Kong Joint Laboratory of Quantum Matter, South China Normal University, Guangzhou 510006, China}

\affiliation{Quantum Science Center of Guangdong-Hong Kong-Macao Greater Bay Area, Shenzhen, China}

\author{Jian-Feng Li}
\affiliation {Key Laboratory of Atomic and Subatomic Structure and Quantum Control (Ministry of Education), Guangdong Basic Research Center of Excellence for Structure and Fundamental Interactions of Matter, School of Physics, South China Normal University, Guangzhou 510006, China} 

\affiliation {Guangdong Provincial Key Laboratory of Quantum Engineering and Quantum Materials, Guangdong-Hong Kong Joint Laboratory of Quantum Matter, South China Normal University, Guangzhou 510006, China}

\author{Yun-Fei Wang}
\email[]{yunfeiwang2014@126.com}
\affiliation {Key Laboratory of Atomic and Subatomic Structure and Quantum Control (Ministry of Education), Guangdong Basic Research Center of Excellence for Structure and Fundamental Interactions of Matter, School of Physics, South China Normal University, Guangzhou 510006, China} 

\affiliation {Guangdong Provincial Key Laboratory of Quantum Engineering and Quantum Materials, Guangdong-Hong Kong Joint Laboratory of Quantum Matter, South China Normal University, Guangzhou 510006, China}

\affiliation{Quantum Science Center of Guangdong-Hong Kong-Macao Greater Bay Area, Shenzhen, China}

\author{Shan-Chao Zhang}
\email[]{sczhang@m.scnu.edu.cn}
\affiliation {Key Laboratory of Atomic and Subatomic Structure and Quantum Control (Ministry of Education), Guangdong Basic Research Center of Excellence for Structure and Fundamental Interactions of Matter, School of Physics, South China Normal University, Guangzhou 510006, China} 

\affiliation {Guangdong Provincial Key Laboratory of Quantum Engineering and Quantum Materials, Guangdong-Hong Kong Joint Laboratory of Quantum Matter, South China Normal University, Guangzhou 510006, China}

\affiliation{Quantum Science Center of Guangdong-Hong Kong-Macao Greater Bay Area, Shenzhen, China}

\author{Hui Yan}
\email[]{yanhui@scnu.edu.cn}
\affiliation {Key Laboratory of Atomic and Subatomic Structure and Quantum Control (Ministry of Education), Guangdong Basic Research Center of Excellence for Structure and Fundamental Interactions of Matter, School of Physics, South China Normal University, Guangzhou 510006, China} 

\affiliation {Guangdong Provincial Key Laboratory of Quantum Engineering and Quantum Materials, Guangdong-Hong Kong Joint Laboratory of Quantum Matter, South China Normal University, Guangzhou 510006, China}

\affiliation{Hefei National Laboratory, Hefei 230088, China}

\author{Shi-Liang Zhu}
\email[]{slzhu@scnu.edu.cn}
\affiliation {Key Laboratory of Atomic and Subatomic Structure and Quantum Control (Ministry of Education), Guangdong Basic Research Center of Excellence for Structure and Fundamental Interactions of Matter, School of Physics, South China Normal University, Guangzhou 510006, China} 

\affiliation {Guangdong Provincial Key Laboratory of Quantum Engineering and Quantum Materials, Guangdong-Hong Kong Joint Laboratory of Quantum Matter, South China Normal University, Guangzhou 510006, China}

\affiliation{Quantum Science Center of Guangdong-Hong Kong-Macao Greater Bay Area, Shenzhen, China}

\affiliation{Hefei National Laboratory, Hefei 230088, China}

\title{\thetitle}

%
\begin{abstract}

Single photons are the flying qubits of choice for distributing entanglement in a quantum internet. Quantum memories embedded in quantum repeaters are crucial to overcome transmission loss and enhance the rate of quantum communication. A multimode memory can further boost the channel capacity. However, benchmarking and building a practical quantum memory that simultaneously optimizes multiple performance metrics poses two key challenges. Here, we introduce quantum interconnect rate to comprehensively quantify quantum memories, and further demonstrate a high-performance quantum memory that simultaneously integrates three essential criteria at once: large multimode capacity, high efficiency, and high fidelity. Operating on 11-dimensional spatial modes, our memory achieves a uniform efficiency exceeding 80\% and qubit storage  fidelities above 99\%, enabling the efficient storage of high-dimensional qudits. Based on these capabilities, we estimate a distribution of $3.56\pm 0.16$ bits of quantum information over a 1000-km repeater link in one minute, highlighting a practical pathway toward scalable quantum interconnects and quantum networks.
\end{abstract}
\maketitle

\textit{Introduction}\textbf{---}
Single photons serve as the fundamental carriers of quantum information in communication networks and quantum interconnects~\cite{kimble2008quantum, wehner2018quantum, azuma2023quantum}. They can encode quantum information in multiple degrees of freedom, such as polarization, time bins, spatial modes, and spectral modes, and can transmit it over long distances with minimal decoherence~\cite{lo2014secure, kuhn2002deterministic, PhysRevLett.70.1895,wang2015quantum}. When entangled with stationary matter qubits, photons enable the distribution of entanglement between distant quantum processors, forming the foundation for distributed quantum computation~\cite{buhrman2003distributed}, non-local quantum sensing networks~\cite{komar2014quantum, nichol2022elementary}, and ultimately a large-scale quantum internet~\cite{kimble2008quantum, wehner2018quantum}. Several important prototypes have already been demonstrated, including high-performance quantum key distribution~\cite{PhysRevLett.134.210801}, metropolitan-scale quantum networks~\cite{liu2024creation, knaut2024entanglement, stolk2024metropolitan}, and proof-of-concept demonstrations of distributed quantum computing~\cite{main2025distributed, wei2025universal}. Together, these advances mark steady progress toward the realization of a global quantum internet in the near future.

However, photon losses in communication channels severely constrain long-distance quantum communication. To overcome this challenge, quantum repeaters (Fig.~\ref{fig0}(a)) are indispensable~\cite{PhysRevLett.81.5932, duan2001long, RevModPhys.83.33}. As the central component of a quantum repeater, a quantum memory synchronizes photonic qubits to enable entanglement swapping and thereby enhance communication rates of quantum key distribution and entanglement distribution~\cite{bhaskar2020experimental, pu2021experimental, PhysRevLett.126.230506, PhysRevLett.129.093604}. Beyond synchronization, quantum memories with multimode storage capability further boost the performance of quantum repeater channels in two key ways. First, the ability to store multiple entangled states enables simultaneous entanglement swapping, providing a direct boost to entanglement distribution rates~{\cite{PhysRevLett.98.060502}}. Second, the storage of high-dimensional photonic qubits and entangled states substantially increases the channel capacity of quantum communication~{\cite{cozzolino2019high, erhard2020advances}}. Multimode quantum memories have been realized across diverse physical platforms, including atomic vapors~\cite{guo2019high, messner2023multiplexed}, cold ensembles~\cite{pu2017experimental, parniak2017wavevector, PhysRevLett.124.210504, yang2025efficient, PhysRevLett.114.050502}, trapped ions~\cite{PhysRevA.106.062617, PhysRevLett.130.213601}, and solid-state systems~\cite{yang2018multiplexed, PhysRevLett.123.080502, chang2025hybrid, PhysRevLett.123.063601, ruskuc2025multiplexed}, where they have been shown to significantly enhance both entanglement distribution rates and channel capacities.

Meanwhile, beyond multimode capacity, the key performance metrics of quantum memory also include efficiency, lifetime, and fidelity~\cite{bussieres2013prospective}. Over the past decade, each of these aspects has witnessed remarkable progress: efficiencies have exceeded 90\%~\cite{wu2025ai}, lifetimes have surpassed one hour~\cite{ma2021one}, fidelities have risen above 99\% ~\cite{wang2019efficient}, and multimode capacities have reached over 1000 modes~\cite{wei2024quantum}. 
However, excelling in any single metric is insufficient for practical quantum interconnects. 
A critical question therefore arises: how to establish a unified and quantitative benchmark that incorporates all key criteria, enabling meaningful evaluation and comparison for real-world applications.
Such a benchmark would motivate the experimental realization of quantum memories that perform strongly across multiple metrics simultaneously.

To address the central challenge of demonstrating and quantifying a high-performance quantum memory for practical quantum interconnects, here, we introduce the \textit{quantum interconnect rate} (QIR) as a comprehensive performance metric and realize a high-performance multimode memory based on a cold atomic ensemble that simultaneously optimizes multiple key performance criteria.
We first provide a theoretical analysis of quantum information transfer in a repeater channel within a fixed time window, showing that multiple performance criteria must be simultaneously satisfied for an optimal quantum memory. 
We then experimentally certify the multimode storage capability of our device by storing single photons in high-order Laguerre–Gaussian (LG) and Bessel–Gaussian  (BG) spatial modes. A uniform storage efficiency exceeding $80\%$ is achieved across 11 modes, while storage fidelities above $99\%$ are verified for 11 groups of photonic qubits, allowing for reliable storage of 11-dimensional photonic qudits with efficiencies above $80\%$ and infidelity below $4\%$.
Using QIR as the comprehensive benchmarking criterion, we estimate $3.56\pm 0.16$ bits of quantum information can be distributed over a 1000-km quantum repeater channel within one minute.
This represents the highest QIR estimated to date compared to previous experiments~\cite{wang2019efficient, PhysRevLett.131.240801, yang2025efficientmultiplex, wei2024quantum, teller2025solid, PhysRevX.14.021018, hartung2024quantum}, paving the way toward practical quantum interconnects in the near future.

\begin{figure}[tp!]
\includegraphics[width=1.0\linewidth]{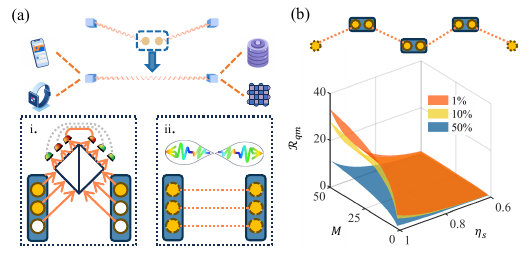}
\caption{{Perfomance of multimode quantum memory}
(a) Scheme of a quantum interconnect based on quantum repeaters. Entanglement swapping within a repeater node (dashed blue box) extends the range of entanglement
between end-nodes (blue cubes) and repeater nodes (yellow spheres).
{The multimode quantum memory enhances the performance of the quantum interconnect in two ways: i. enabling parallel entanglement swapping through multiplexing, and ii. increasing the channel capacity via high-dimensional storage.}
(b) Comprehensive performance of a high-dimensional quantum memory. The performance is quantified for a 1000-km, 2-layer quantum repeater channel. The QIR, $\mathcal{R}_{qm}$, is plotted as a function of the number of modes $M$ and the storage efficiency $\eta_s$, under different levels of depolarizing noise ($p_n = 1\%, 10\%, 50\%$). The other parameters are set as follows: $\eta_d=1$, $p=1$, and $L_{\text{att}}=22~\mathrm{km}$.   
}\label{fig0}
\end{figure}

\textit{Quantum interconnect rates}\textbf{---}
Quantum memory play a central role in quantum repeater protocols, serving two key functions. First, they enhance secure key rates and accelerate entanglement distribution by synchronizing photons for entanglement swapping. Second, their multimode capacity further boosts performance, supporting either multiplexed entanglement swapping or high-capacity communication channels. To benchmark memory performance in this context, we introduce \textit{QIR}:
\begin{equation}
\label{eqR}
 \mathcal{R} = C/T,
\end{equation}
where $C$ is the information capacity of a quantum channel for transmitting photonic qudits and $T$ is the corresponding photon distribution time. Thus, $\mathcal{R}$ represents the rate at which quantum information can be distributed across long distances.
{Since our focus is on the performance of quantum memories, we assume near-ideal entangled-state sources and communication channels. In particular, the photonic states are treated as pure, and decoherence in the transmission channel is neglected. We further assume that the transmission channel supports an effectively infinite number of modes, such that the multimode capacity is primarily limited by the quantum memory rather than by the channel itself.}

Encoding photonic qudits in a large Hilbert space increases the information per photon by spanning spatial, temporal, or spectral degrees of freedom significantly, and thus the achievable entanglement distribution rate.
Likewise, storing photonic qudits enhances the capacity of a quantum memory, which can be quantified as~\cite{PhysRevLett.118.050501}:
\begin{equation}
\label{eqC}
C_{qm}=\max_{p(x)}\sum_{y=1}^{M}\sum_{x=1}^{M}p(y|x)p(x)\log_2\left(\frac{p(y|x)}{\sum_{x'=1}^{M}p(y|x')p(x')}\right),
\end{equation}
where $M$ is the number of supported modes, $x$ and $y$ denote the input and retrieved states, $p(x)$ is the input distribution, and $p(y|x)$ is the conditional retrieval probability.
In the ideal case, the maximum capacity is $\log_2{M}$. In practice, imperfect storage reduces $C_{qm}$, as well as the storage fidelity. For example, if a depolarizing noise $p_n$ acts on the stored state $\rho_{in}$, the retrieved state becomes $\rho_{out}=(1-p_n)\rho_{in}+({p_n}/{M})I$, resulting in crosstalk error $p(y|x)=p_n/M$ (for $y\neq x$), and storage fidelity $F_s=1-(M-1)p_n/M$. Such crosstalk error may also reduce the channel capacity \cite{SupM}.

Long-distance quantum networking requires the distribution of entanglement between remote nodes. For an n-nesting-level quantum repeater, the expected entanglement distribution time is given by~\cite{PhysRevLett.98.190503, RevModPhys.83.33}
\begin{equation}
\label{eqTtot}
\begin{aligned}
T_{tot}=3^{n+1}\frac{L_0}{c}\frac{\prod_{k=1}^n\left(2^k-(2^k-1)\eta\right)}{Np\eta_{L_0}\eta_d\eta^4},
\end{aligned}
\end{equation}
where $N$ is the number of multiplexing modes, $p$ is the probability of generating an entangled photon pair, $\eta=\eta_s\eta_d$ denotes the overall efficiency which includes the storage efficiency ($\eta_s$) and detection efficiency ($\eta_d$). The transmission efficiency for a segment of $L_0$ is  $\eta_{L_0}=e^{-L_0/(2L_{att})}$,  where $c$ is the speed of light in fiber and $L_{att}$ is the fiber attenuation length determined by the fiber losses at photon wavelength. This expression incorporates the effects of both memory efficiency and parallel multiplexing strategies~\cite{PhysRevLett.98.190503, PhysRevLett.98.060502}. 

For qudit-based repeater networks, the QIR is given by $\mathcal{R}_{qm} = C_{qm}/T_{tot}$. This metric provides a unified benchmark that captures all key memory properties: multimode and multiplexing capacity, fidelity, and efficiency.
As a representative scenario, we consider a 1000-km quantum repeater channel comprising four segments in a two-nesting-level configuration~\cite{PhysRevLett.98.190503}. As illustrated in Fig.~\ref{fig0}(b), achieving optimal performance requires simultaneously realization of high storage efficiency, large multimode capacity, and high-fidelity operation. Guided by the metric $\mathcal{R}_{qm}$, we experimentally demonstrate a quantum memory that achieves high performance across these three aspects in the following sections.

\begin{figure}[]
\includegraphics[width=1.0\linewidth]{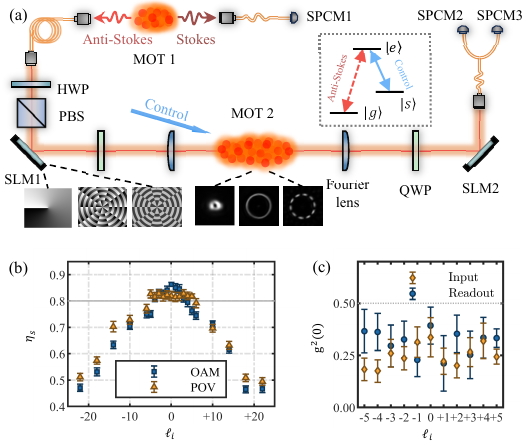}
\caption{{Efficient quantum memory for single photons in spatial modes.}
(a) Experimental setup. Two $^{85}$Rb atomic ensembles are employed. 
After transmission through 200 m of fiber, the anti-Stokes photons are encoded into
OAM or POV modes, using SLM 1. The encoded photons are stored in MOT 2 , retrieved by programmable control pulses, decoded with SLM 2, and guided to SPCMs 2 and 3 via a fiber beam splitter.
The inset shows three examples of phase patterns on SLM 1 and their corresponding beam profiles at the atomic ensemble, including an OAM mode with $\ell=+1$, a POV mode with $\ell=+5$, and a superposed POV mode $\ell_1+\ell_2$ where $\ell_1=+5$ and $\ell_2=-5$.
The energy levels involved in the EIT process are shown in the dashed box: the ground state $\ket{g}=\ket{5S_{1/2},~F=2}$, the storage state $\ket{s}=\ket{5S_{1/2},~F=3}$, and the excited state $\ket{e}=\ket{5P_{1/2},~F=3}$. 
(b) Storage efficiency. Single photons encoded with OAM (blue) and POV (yellow) modes are stored and retrieved to quantify the storage efficiency. {$1.8\times10^5$ experimental cycles are implemented to determine storage efficiency of each mode.}
(c) Single-photon autocorrelation. The second-order correlation function $g^{(2)}(0)$ is plotted as a function of POV mode order. Error bars indicate one standard deviation.
}
\label{fig1}
\end{figure}

\textit{Efficient multimode quantum memory}\textbf{---}
The experimental scheme is shown in Fig.~\ref{fig1}. It consists of two $^{85}$Rb ensembles confined in magneto-optical traps (MOT 1 and MOT 2), with all atoms initially prepared in the ground state $\ket{g}$. 
The correlated Stokes–anti-Stokes photon pairs are generatated in MOT~1 via a spontaneous four-wave mixing process~\cite{PhysRevLett.106.033601}.
We employ photon-shaping techniques to obtain Gaussian-shaped biphoton waveforms and match their spectral–temporal properties to the memory~\cite{li2019generation}. The detection of a Stokes photon by a single-photon counting module (SPCM 1) heralds its correlated anti-Stokes photon transmitted to MOT 2.
To encode qudit states, the anti-Stokes photons are modulated by a spatial light modulator (SLM 1), which imprints phase patterns corresponding to LG or BG modes, as illustrated in the insets of Fig.~\ref{fig1}(a). After passing through a Fourier lens, the anti-Stokes photon is transformed into orbital angular momentum (OAM) or perfect optical vortex (POV) states and stored as a collective spin wave in MOT 2 via electromagnetically induced transparency (EIT) using a programmable control beam. After a storage period, the photonic qudit is retrieved, decoded by SLM 2, and detected by SPCMs 2 and 3.
The memory performance of MOT 2 is characterized using a weak probe beam. The optical depth (OD) is measured to be 280, and the memory efficiency reaches $87.6\%$, with a memory lifetime of $28.0~\mu\text{s}$~\cite{SupM}. {The storage efficiency can be further improved by increasing OD of the ensemble.~\cite{PhysRevLett.120.183602, wang2019efficient}}

Multimode storage capability is crucial for increasing the channel capacity of quantum repeater networks. In our system, the storage efficiency relies on a cigar-shaped ensemble with a large OD along the photon propagation direction. This configuration makes transverse spatial modes a natural choice for extending multimode storage. In the spatial domain, photons can carry both spin angular momentum (SAM) and OAM~\cite{PhysRevA.45.8185, shen2019optical}. While SAM offers only two dimensions corresponding to polarization states, OAM supports in principle an infinite set of modes, enabling the construction of high-dimensional states with broad applications in optical communications~\cite{willner2015optical}.
Here, we encode the anti-Stokes photons into either OAM or POV states, and characterize the storage efficiency by counting the retrieved photons. 
The measured storage efficiencies for orbital orders from $\ell=-22$ to $+22$ are presented in Fig.~\ref{fig1}(b). For OAM modes, the maximum efficiency of $86.3\pm0.8\%$ occurs at $\ell=0$, corresponding to the Gaussian mode. The efficiency decreases with increasing $|\ell|$, reaching values around $50\%$, which still surpasses the quantum no-cloning limit without post-selection. This behavior arises because the beam waist of LG modes scales as $\sqrt{|\ell|}$, leading to a reduced overlap with the effective OD of the atomic ensemble. For POV modes, the storage efficiency remains nearly constant, averaging $82.2\%$ across modes from $\ell=-5$ to $+5$. At higher orbital orders, however, the efficiency also gradually decreases.

A key advantage of POV states is that their core size and intensity distribution remain nearly independent of the orbital order, leading to a uniform OD and storage efficiency across a range of modes~\cite{ostrovsky2013generation}. This property is particularly beneficial for realizing high-dimensional quantum memories. Although POV states are less stable in free-space propagation, they can be directly generated from LG beams using an axicon and Fourier lenses~\cite{jabir2016generation, liu2017generation}. This makes it feasible to integrate a quantum memory for POV states into high-dimensional communication channels based on LG modes in free space or OAM fibers~\cite{wang2016advances}.
The measured second-order autocorrelation, 
$g^{(2)}(0)$, remains below 0.5 both before and after storage across 11 POV modes, confirming that the single-photon character is preserved when an anti-Stokes photon is encoded in POV states and retrieved from the quantum memory. In the following sections, we focus on these 11-dimensional POV states and measure the storage fidelity of qubits and qudits.

\begin{figure}[t]
\includegraphics[width=1.0\linewidth]{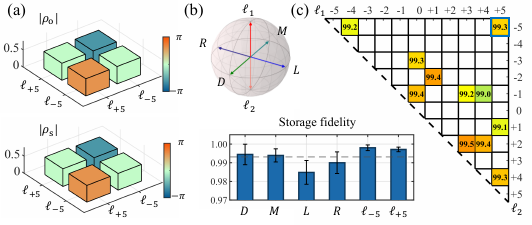}
\caption{{Storage fidelity of photonic qubits.}
(a) Density matrix of a photonic qubit before and after storage. Two POV modes, $\ell_1=+5$ and $\ell_2=-5$, form a qubit. The density matrix of the state $\ket{\varphi}=(\ket{\ell_1} + i\ket{\ell_2})/\sqrt{2}$ is reconstructed as $\rho_o$ (before storage) and $\rho_s$ (after storage), respectively.
(b) Storage fidelity within a qubit subspace. Six states are tested in the two-dimensional subspace defined by $\ket{\ell_1}$ and $\ket{\ell_2}$: $D=(\ket{\ell_1}+\ket{\ell_2})/\sqrt{2}$, $M=(\ket{\ell_1}-\ket{\ell_2})/\sqrt{2}$, $L=(\ket{\ell_1}+i\ket{\ell_2})/\sqrt{2}$, $R=(\ket{\ell_1}-i\ket{\ell_2})/\sqrt{2}$, $\ket{\ell_1}$, and $\ket{\ell_2}$. These states are represented on the Bloch sphere. The storage fidelity for each state is shown below, with the average fidelity $\bar{F}=99.3\%$ indicated by the dashed line. 
(c) Average storage fidelity of qubits. From 55 possible qubit pairs, 11 are randomly selected for fidelity measurement. The average fidelity for the six states of each selected qubit is displayed in the corresponding cell of the table (in percentage).
}
\label{fig3}
\end{figure}

\textit{High-fidelity storages}\textbf{---}
To demonstrate a practical application of our quantum memory, we first characterize its performance for storing photonic qubits. A two-dimensional Hilbert subspace is defined using two orbital orders, $\ell_1$ and $\ell_2$, selected from the 11 available POV states. We employ quantum state tomography with maximum-likelihood estimation to reconstruct the density matrices of the photonic qubit states before ($\rho_o$) and after ($\rho_s$) storage. An example of these reconstructed density matrices is presented in Fig.~\ref{fig3}(a). The storage fidelity is then calculated using the formula $F=\left[\mathrm{Tr}\left(\sqrt{\sqrt{\rho_o}\rho_s\sqrt{\rho_o}}\right)\right]^2$.

In each selected subspace, six qubit states are prepared to evaluate the storage fidelity, as shown in Fig.~\ref{fig3}(b). Experimentally, 11 out of the 55 possible qubit subspaces are randomly chosen for characterization, and the average storage fidelities for each subspace are summarized in Fig.~\ref{fig3}(c). In all measured cases, the storage fidelity exceeds $97\%$, with an overall average of $\bar{F} = 99.3 \pm 0.4\%$.

Next, we verify the high-dimensional storage capability using an 11-dimensional photonic qudit.
By coherently superposing all orbital orders, the heralded qudit state is expressed as $\ket{\Psi}=\frac{1}{\sqrt{11}}\sum_{\ell_i=-5}^{+5}\ket{\ell_i}$.

In practice, the correlated counts between the Stokes and anti-Stokes photons include accidental coincidences and double-pair excitations arising from the lossy detection channel.
The heralded qudit state can therefore be described as
\begin{equation}
\label{eqrhoe}
\rho_e = p_0\rho_0 + p_1\rho_1 + p_2\rho_2,
\end{equation}
where $p_0$, $p_1$, and $p_2$ denote the probabilities of detecting zero, one, or two anti-Stokes photons, corresponding to the density matrices $\rho_0$, $\rho_1$, and $\rho_2$, respectively. Contributions from higher-order excitations are negligible in the four-wave mixing process and are therefore omitted. Thus, the state fidelity is defined as $F=\bra{\Psi}\rho_e\ket{\Psi}$. Since the population across the modes of the qudit state is not uniformly distributed, we derive a lower bound for the qudit-state fidelity using the measured mode populations, $q_f$, given by
\begin{equation}
\label{eqFide}
F \geq \frac{p_1 q_f}{T (p_1 + 2 p_2)} \left( \sum_i \sqrt{t'_i / 11} \right)^2,
\end{equation}
where $T=\sum_i\mathcal{h}^{(i)}/11$, $t'_i=\mathcal{h}^{(i)}/11T$, and $\mathcal{h}^{(i)}$ is the heralding rate of the $i$-th mode of the correlated anti-Stokes photons~\cite{pu2018experimental}.

To determine $p_0$, $p_1$, and $p_2$, we measure the heralding rates of the anti-Stokes photon in each spatial mode, $\mathcal{h}^{(i)}_o$ ($\mathcal{h}^{(i)}_s$), as well as the qudit population distributions, $q^{(i)}_o$ ($q^{(i)}_s$), before (after) storage in the quantum memory, as shown in Fig.~\ref{fig4}.
The measured storage efficiency of the qudit state is $84.3 \pm 2.4\%$, and the autocorrelation is $0.35 \pm 0.12$ ($0.37\pm 0.15$) before (after) storage.
The fidelities of the photonic qudit before and after storage are $F_o \geq (87.2 \pm 0.5)\%$ and $F_s \geq (84.1 \pm 0.5)\%$, respectively. The small decay of only $3.6\%$ indicates that the 11-dimensional qudit is well preserved during quantum storage.
Details of the measurement procedure are provided in the Supplemental Material~\cite{SupM} and in Refs.~\cite{pu2018experimental, PhysRevLett.124.240504}.

\begin{figure}[t]
\includegraphics[width=1.0\linewidth]{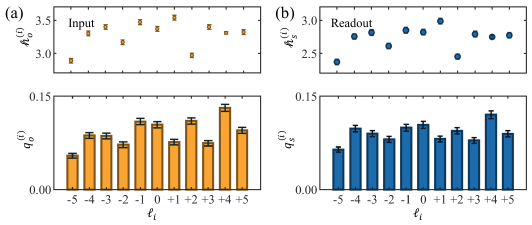}
\caption{{Characterization of photonic qudits.}
Qudit states are characterized before storage ({a}) and after storage ({b}). For both the input and retrieved states, we measure the heralding rate $h_o^{(i)}$ (before) and $h_s^{(i)}$ (after), as well as the population distribution $q_o^{(i)}$ (before) and $q_s^{(i)}$ (after) for each POV mode.}
\label{fig4}
\end{figure} 

\textit{Discussion and conclusion}\textbf{---}
Finally, we benchmark our quantum memory using the performance metric $\mathcal{R}_{qm}$ within the 1000-km repeater architecture depicted in Fig.~\ref{fig0}(b). The evaluation assumes the storage of 11 spatial modes with an average storage efficiency of $82.2\pm 0.5\%$ and an average qubit storage fidelity of $99.3\pm 0.4\%$. We also assume ideal detection efficiency ($\eta_d=1$) and perfect entanglement swapping operations ($p_s=1$).
For our absorptive-type memory, we consider a deterministic photon-pair source with an emission probability of $p=0.7$~\cite{ding2025high}. 
{Under these conditions, we estimate the channel capacity $C_{qm}=3.32\pm 0.07$ and distribution time $T_{tot}=56.0\pm 2.2$s for our 11-mode quantum memory, corresponding to a QIR of $3.56\pm 0.16$ bits per minute over the 1000-km channel~\cite{SupM}. The error bars are determined by assuming Poissonian photon-counting statistics and propagating the uncertainties through numerical Monte Carlo simulations.}
Notably, the intrinsic coherence time of atomic qubits has already reached the minute scale~\cite{wang2021single, tian2024extending, young2020half}, which positions our system as a practical platform for long-distance quantum interconnects between remote processors. We emphasize that further advances in deterministic bright entangled-photon sources and low-noise optical links will be particularly critical for realizing scalable and deployable quantum network infrastructures.

In summary, we address a central challenge in realizing a high-performance quantum memory suitable for practical quantum interconnects.
We have demonstrated an efficient quantum memory for storing single photons in spatial modes. The memory is capable of storing over 30 distinct modes with a storage efficiency exceeding 50\%. Notably, it achieves a uniform efficiency above 80\% across 11 photonic POV states.
{Encoded in this 11-dimensional space, photonic qudit states can be stored with high and uniform efficiency. This preserves the photon signal-to-noise ratio, suppresses depolarizing noise during storage, and mitigates amplitude damping across modes, thereby enabling high-fidelity storage of both qubits and qudits.}
To benchmark the overall performance of our quantum memory, the QIR is introduced as a comprehensive metric that simultaneously incorporates memory lifetime, storage efficiency, fidelity, and multimode capacity. 
Additional performance criteria can also be naturally incorporated into the QIR framework, including the operating wavelength through $L_{att}$ and memory bandwidth through frequency and temporal multiplexing capabilities.

Several avenues for further improvement are within reach. The memory lifetime could be extended to the millisecond scale by applying dynamical decoupling pulses to freeze the spin-wave~\cite{PhysRevA.93.063819, PRXQuantum.2.040307}, and potentially to the second scale via optical trapping of the atomic ensemble~\cite{PhysRevA.87.031801}. Quantum memory arrays could be realized by confining micro-ensembles in optical tweezers integrated with cavities~\cite{wang2020preparation, shaw2025cavity}. Finally, the integration of multiple degrees of freedom into the present platform represents a promising route for substantially increasing the channel capacity~\cite{parigi2015storage}.

\textit{Acknowledgments}\textbf{---}This work is supported by the National Key R\&D Program of China (Grants Nos.~2020YFA0309500, 2022YFA1405300), the National Natural Science Foundation of China (Grants Nos.~12404407, 12534011, 12322408, 12225405, 62371198), Quantum Science and Technology-National Science and Technology Major Project (Grant No.~2021ZD0301700), Guangdong Provincial Quantum Science Strategic Initiative (Grant Nos.~GDZX2504005, GDZX2503007, GDZX2404001, GDZX2404003, GDZX2304002), and Guangdong Basic and Applied Basic Research Foundation (Grant Nos.~2026A1515030019, 2025A1515011684, 2022A1515110921).

\textit{Data availability}\textbf{---}The data that support the findings of this article are not publicly available. The data are available from the authors upon reasonable request.

%


\newpage

\begin{widetext}


\maketitle

\section{Supplemental Material for \\ A high-performance quantum memory for quantum interconnects}

\setcounter{equation}{0}
\setcounter{figure}{0}
\setcounter{table}{0}
\setcounter{page}{1}
\setcounter{section}{0}
\makeatletter
\renewcommand{\theequation}{S\arabic{equation}}
\renewcommand{\thefigure}{S\arabic{figure}}
\renewcommand{\thetable}{S\arabic{table}}
\FloatBarrier

\section{I. Details of Experiment}

The experiment consists of two $^{85}$Rb atomic ensembles confined in magneto-optical traps (MOT~1 and MOT~2), as illustrated in Fig.~\ref{figsetup}. Heralded anti-Stokes photons generated from MOT~1 are collected through a 200~m polarization-maintaining single-mode fiber (SMF) and transmitted to MOT~2 for storage. The two MOT systems are located on separate optical tables to minimize optical and magnetic cross-interference.
For MOT~1, the typical optical depth (OD) of the atomic ensemble is about 100, with a magnetic field gradient of approximately 16~G/cm. Each of the six trapping beams has a power of 20~mW and a diameter of 2~cm. The repumping beams have a total power of 70~mW and the same beam diameter, with central dark lines of 1.0~mm width to suppress unwanted scattering and heating. A 4-f imaging system projects these dark lines precisely onto the atomic ensemble, ensuring uniform atom density and high OD in the trapping region.
For MOT~2, the ensemble exhibits a typical OD of 280, and the magnetic field gradient is about 9~G/cm. Each trapping beam has a power of 50~mW and a diameter of 4~cm. The two repumping beams have a combined power of 80~mW, sharing the same diameter as the trapping beams. Two 1.5~mm-wide dark lines are placed along the centers of the repumping beams, and two 4-f imaging systems project these lines onto the ensemble to optimize atomic density and optical homogeneity.

\begin{figure*}[h]
\includegraphics[width=1\linewidth]{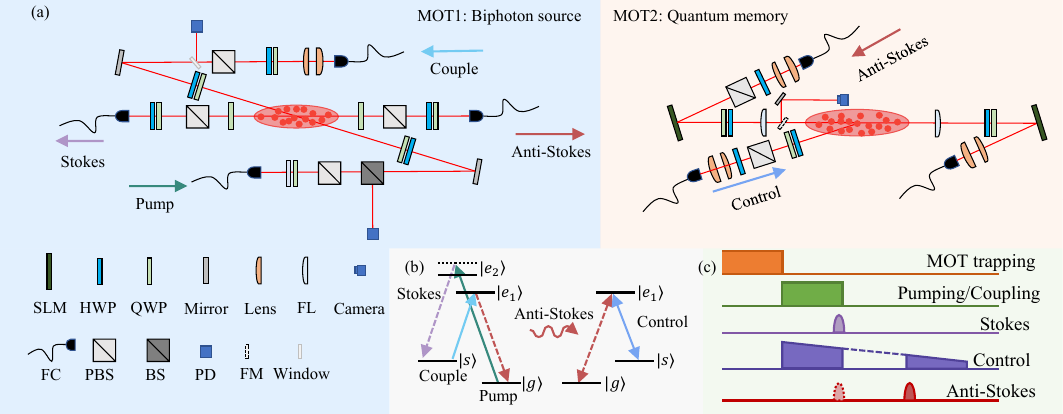}
\caption{\textbf{Experiment setup}. 
\textbf{a}, Schematic of the experiment. Two $^{85}$Rb atomic ensembles serve as the biphoton source and the quantum memory, respectively. The intensities of the pump and coupling beams are monitored by photodetectors (PDs) and actively stabilized using half-wave plates (HWPs), quarter-wave plates (QWPs), and polarizing beam splitters (PBSs) for polarization control. The anti-Stokes photons are projected onto spatial light modulators (SLMs) for encoding and decoding qudit states. A flip mirror (FM) directs the beam toward a Fourier lens (FL) and an auxiliary camera to monitor the spatial mode. 
\textbf{b}, Relevant energy levels of $^{85}$Rb and the corresponding optical transitions for the pump, coupling, and photon fields. 
\textbf{c}, Experimental timing sequence. During the measurement window, the trapping beams are turned off. Upon detection of a Stokes photon, the control beam is switched off and on to store and retrieve the correlated anti-Stokes photon. The control beam amplitude is linearly ramped to compensate for the decreasing optical depth and to maximize the storage efficiency.
}\label{figsetup}
\end{figure*} 

Heralded single photons are generated from a time–frequency entangled photon-pair (biphoton) source based on spontaneous four-wave mixing (SFWM) in MOT~1. Under the action of counter-propagating circularly polarized pump (780~nm, $\sigma^{-}$, 3~$\mu$W) and coupling (795~nm, $\sigma^{+}$ polarization, 6.5~mW) laser beams, phase-matched Stokes (780~nm, $\sigma^{-}$) and anti-Stokes (795~nm, $\sigma^{+}$) photons are generated in opposite directions. The pump laser is blue-detuned by 80~MHz from the $\ket{g}=\ket{5S_{1/2},F=2}\rightarrow\ket{e_2}=\ket{5P_{3/2},F=3}$ transition and focused at the center of MOT~1 with a $1/e^2$ beam diameter of 0.4~mm. The coupling laser, resonant with the  $\ket{s}=\ket{5S_{1/2}, F=3}$ to $\ket{e_1}=\ket{5P_{1/2}, F=3}$ transition, is a collimated Gaussian beam with a $1/e^2$ diameter of 5~mm. The angle between the Stokes/anti-Stokes propagation axis and the pump/coupling axis is set to $2.7^{\circ}$, ensuring spatial separation of the weak Stokes and anti-Stokes fields from the strong pump and coupling fields.
The paired photons are coupled into two single-mode fibers (SMFs) with mirrored spatial modes and overlapping foci at the MOT~1 center, each with a $1/e^2$ mode-field diameter of 0.24~mm. The biphoton coherence time, measured from the coincidence histogram, has a full width at half maximum (FWHM) of approximately 350~ns. Two Fabry–Perot etalon filters (bandwidth 210~MHz, isolation ratio 55~dB) are used to suppress residual background photons in both Stokes and anti-Stokes channels. The photons are detected by three single-photon counting modules (SPCMs), and coincidence events are recorded using a time-to-digital converter (Fast Comtec P7888) with a 1~ns time bin. The measured biphoton coincidence rate is 12~pairs/s.
Considering the fiber–fiber coupling efficiency ($85\%$), quantum-memory system transmission ($40\%$), filter transmission ($50\%$), SPCM detection efficiency ($60\%$), and experimental duty cycle ($3\%$), the intrinsic photon-pair generation rate at the source is estimated to be approximately $4\times10^{3}$~pairs/s. Detection of a Stokes photon heralds the presence of a single anti-Stokes photon in a well-defined temporal mode. The measured heralding efficiency is $3.5\%$, corresponding to $17.1\%$ at the source after correcting for optical and detection losses. These losses affect only the overall count rate, not the intrinsic performance of the quantum memory. The optimal Gaussian-like single-photon temporal waveform is obtained following the optimization procedure described in Refs.~\cite{wang2019efficient}.

The cold {$^{85}$Rb} atomic ensemble in MOT~2 serves as the quantum memory. The ensemble has dimensions of $4\times4\times30~\text{mm}^3$ and an optical depth (OD) of 280. The atomic temperature is approximately $125~\mu\text{K}$. The heralded anti-Stokes signal beam is focused at the center of MOT~2 with a beam waist radius of $30~\mu\text{m}$.
During each experimental cycle, the quadrupole magnetic field of MOT~2 is switched off prior to the 0.3~ms memory operation window. At the beginning of this window, a collimated control laser with a $1/e^2$ beam diameter of 6~mm is turned on. Upon detection of a Stokes photon, the heralded anti-Stokes photon enters the atomic ensemble. The control field is then rapidly switched off, with a 30~ns fall time, using an acousto-optic modulator (AOM) driven by a digital waveform generator (Rigol DSG815). This process maps the flying anti-Stokes photon into a long-lived collective atomic excitation (spin wave). After a programmable storage time, the control laser is turned back on to retrieve the stored photonic state.

Stokes photons are detected by SPCM~1, which also triggers the experimental sequence for storing the corresponding anti-Stokes photons. The MOTs are reloaded at a repetition rate of 100~Hz, and each experimental cycle occupies a 300~$\mu$s time window. The temporal length of the anti-Stokes photon wave packet is measured to be approximately 800~ns, and an identical time window is used to record the correlated anti-Stokes photons.

The second-order autocorrelation function is calculated as,
\begin{equation}
\begin{aligned}
g_c^{2}(0)=N_1N_{2,3}/(N_2N_3),
\end{aligned}
\end{equation}
where $N_1$ denotes the Stokes photon counts recorded by SPCM~1, $N_2$ ($N_3$) are the anti-Stokes photon counts from SPCM~2 (SPCM~3) within the correlation window, and $N_{2,3}$ represents the coincidence counts of simultaneous detections on SPCM~2 and SPCM~3. Counts registered outside the correlation window originate from background noise, which allows us to estimate accidental coincidences and subtract them from the raw data when necessary.

Each experimental run lasts for 300~s, during which approximately $10^5$ Stokes photons are detected. The heralding rate of the anti-Stokes photons is about $2.5–3.5\%$, calculated as
\begin{equation}
\begin{aligned}
\mathcal{h}^{(i)}=(N_2+N_3)/N_1,
\end{aligned}
\end{equation}
The storage efficiency is then obtained as
\begin{equation}
\begin{aligned}
\eta_s=\mathcal{h}^{(i)}_s/\mathcal{h}^{(i)}_o,
\end{aligned}
\end{equation}
and each measurement is repeated three times to determine the statistical uncertainty, without subtracting accidental coincidences.

{In practice to determine the storage efficiency and autocorrelation of each spatial mode, $1.8\times10^5$ experimental cycles are implemented in 30 minutes, including three measurement runs without storage and three runs with photon storage in the atomic ensemble. Measurements without and with storage are performed alternately to compensate for long-term fluctuations in the photon-pair generation rate. }

For the qubit storage experiments, two random numbers are generated to select a specific two-dimensional subspace from the available spatial modes. The qubit density matrix is reconstructed via quantum state tomography using the maximum-likelihood estimation method. Four measurement bases are employed for complete state reconstruction, which are $\ell_1$, $\ell_2$, $D$, and $L$.
To estimate the lower bound of the qudit-state fidelity, the heralding rate of each mode is determined after subtracting accidental coincidences, which enables accurate estimation of the double-excitation probability.

\section{II. Calibration of the experiment}


\begin{figure}[b]
\includegraphics[width=1.0\linewidth]{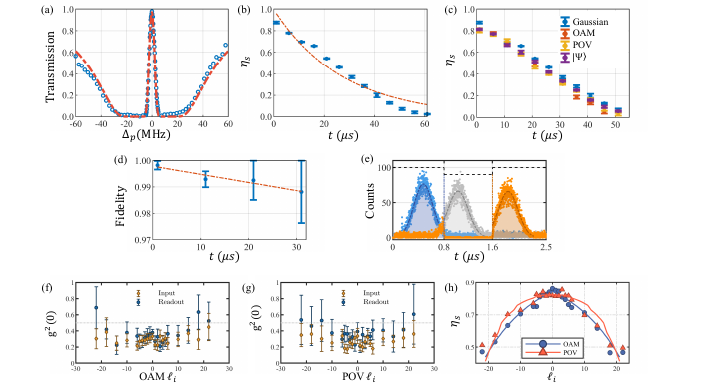}
\caption{{Experiment Calibration}. 
{(a)} EIT transmission spectrum. {The blue circles represent the measured EIT transmission profile of the ensemble, and the red dashed line is the fitted theoretical curve.} The measured OD with a Gaussian probe beam is 280.
{(b)}, Memory efficiency as a function of storage time. The blue dots represent the measured storage efficiencies using weak probe pulses in Gaussian mode, each with an average photon number of about 0.5 per pulse. The dashed line shows an exponential fit to the data. 
{(c)}, Storage lifetime for different spatial modes. Four cases are given to verify the storage time, including weak probe pulses in Guassian (blue), {OAM (red, $l=+5$), POV (orange, $l=+5$)} and 11-dimensional qudit (violet) mode. The average storage lifetime is approximately $28.0~\mu s$.
{(d)}, Qubit fidelity versus storage time. The probe pulses are encoded in the qubit state $\ket{\varphi}=(\ket{\ell_{+5}}+\ket{\ell_{-5}})/\sqrt{2}$, and the state fidelity is reconstructed via quantum state tomography at varying storage times. Error bars are derived from Monte Carlo simulations assuming Poissonian photon-count statistics.
{(e)}, Temporal waveforms of heralded single photons. The single photon is projected onto the POV mode $\ell_{+5}$, showing three representative cases: generated (blue), EIT-delayed (gray), and retrieved (orange).
{The second-order correlation function $g^{(2)}(0)$ is plotted as a function of OAM (f) and POV (g) mode order.  Error bars indicate statistical error of triple measurements when anti-Stokes photons are passing through (yellow) or stored by (blue) the quantum memory.
{(h)}, Storage efficiency $\eta_s$ of OAM (blue) and POV modes (red) as a function of mode index $\ell_i$ with experiments (scatters) and numerical simulation (lines).The simulation considers the overlapping of spatial beams and atomic ensemble}
}\label{figslife}
\end{figure} 

Prior to the single-photon storage experiment, we calibrated the quantum memory in MOT~2 using a weak coherent probe beam of which the frequency matched that of the anti-Stokes photons. We first measured the optical depth (OD) of the atomic ensemble to confirm that the insertion of SLM~1 and SLM~2 did not perturb the optical paths. By scanning the probe detuning, we obtained the EIT transmission spectrum and extracted the effective OD for different spatial modes, as shown in Fig.~\ref{figslife}(a).
Next, we characterized the storage efficiency and lifetime of the quantum memory. A weak coherent probe pulse was modulated by an electro-optic modulator to generate a Gaussian waveform with a full width at half maximum (FWHM) of $350~\text{ns}$, matching the temporal profile of the heralded single photons. For MOT~2, the storage efficiency for coherent light in the Gaussian mode reached $87.6\%$ at a $1/e$ storage time of $800~\text{ns}$, and the storage lifetime was measured to be $29~\mu\text{s}$, as shown in Fig.~\ref{figslife}(b).

{The storage efficiency is mainly limited by the finite OD of our current atomic ensemble, and it can reach more than $90\%$ if OD is larger than 500~\cite{wang2019efficient}. The storage time in current experiment is mainly limited by dephasing of the stored spin-wave caused by the atomic motion and small residual gradient of magnetic field. Several methods could be applied to extend storage time. First one is to apply dynamical decoupling pulses to freeze the spin-wave~\cite{PhysRevA.93.063819, PRXQuantum.2.040307}, because it can compensate the kinetic vector of spin-wave as well as transfer the stored collective excitation to a pair of clock states. The storage time can be extended to millisecond scale with this method, but it is still limited by diffusion in free space as well as free-falling by gravity. So the second method is to trap atoms in optical lattice which confines the atomic motion alongside the kinetic vector direction~\cite{PhysRevA.87.031801}. Such method could extend storage time to second-scale but the optical lattice should be carefully designed because the length of our atomic ensemble is 3-centimeter, which demands optical lattices at same scale. A possible design is to combine a three-dimensional box trap with a blue-detuned grid trap~\cite{PhysRevA.95.013632, PhysRevLett.123.230501}. }

We repeated the measurements for higher-order spatial modes, obtaining similar storage lifetimes of approximately $28~\mu\text{s}$ across all modes (Fig.~\ref{figslife}(c)). {The OAM and POV modes we used here have the same orbital order $\ell=+5$, and beam diameters are approximately $150~\mu m$. Therefore, they experience nearly identical effective optical depths in the ensemble, resulting in similar storage efficiencies. Spatial modes possess azimuthally varying phase structures, and atomic motion may degrade the corresponding transverse spin-wave coherence. The characteristic coherence length of the spatial mode can be estimated as $\lambda_{POV}=\pi d/|\ell|$, yielding $\lambda_{POV}\approx95\mu m$ for $d=150~\mu m$ and $|\ell|=5$. By comparison, the spin-wave coherence length is $\lambda_{sw}=795nm/2^\circ \approx 23\mu m$, determined by the angle between the probe and control beams. Since $\lambda_{POV}\gg \lambda_{sw}$, the memory lifetime is mainly limited by spin-wave decoherence, leading to nearly identical storage lifetimes for different spatial modes in the present system.}

We further measured the storage lifetime of a qubit state, $\ket{\varphi}$, reconstructed via quantum state tomography. As shown in Fig.~\ref{figslife}(d), the fitted $1/e$ lifetime was $3.0~\text{ms}$. After completing the calibration, we performed the single-photon storage experiment. Figure~\ref{figslife}(e) shows an example of the temporal waveforms of the input (blue), EIT-delayed (gray), and retrieved (orange) heralded anti-Stokes photons at the optimized storage efficiency. Photon coincidences were accumulated over 1200~s, and the solid curves represent theoretical fits. The black dashed line marks the switching-off moment of the control beam, and the storage period, bounded by the blue and orange dashed lines, corresponds to a one-pulse delay of $0.8~\mu\text{s}$.
{In Fig.~\ref{figslife}(f) and (g), we present the second-order correlation function $g^{(2)}(0)$ of all measured OAM modes and POV modes respectively. After encoded in spatial modes, the anti-Stokes photons can still maintain their single photon property since the correlation remains below 0.5 as indicated by the yellow spots. After retrieved from quantum memory, $g^{(2)}(0)$ becomes larger than 0.5 when the spatial mode order is higher (blue spots). We attribute it to the lower storage efficiency of high-order spatial modes so that photon count portion from the multiple excitations and background noise during storage process becomes larger. }

{The ideal POV modes have $\ell$-independent beam diameter, which is derived from the Fourier transformation of a Bessel function. However, we employ Bessel–Gaussian beams rather than ideal Bessel modes, resulting in a mode-dependent waist,~\cite{Pinnell19, PhysRevLett.131.240801}
\begin{equation}
\label{eqABeamW}
\begin{aligned}
\omega_{\ell}=\omega_{0}\sqrt{|\ell|+1}+r_r\sqrt{1+\frac{I_{\ell+1}(r_r^2/\omega_0^2)}{I_{\ell}(r_r^2/\omega_0^2)}},   
\end{aligned}
\end{equation}
where $\omega_0$ is the gaussian beam waist at Fourier plane, $r_r$ is the ideal POV radius, and $I_{\ell}$ is the modified Bessel function. The front half of above function, $\omega_0\sqrt{|\ell|+1}$, is the same beam waist of OAM modes, and it dominates the POV waist when $|\ell|$ is large. Consequently, higher-order POV modes exhibit larger beam sizes, leading to reduced effective overlap with the atomic ensemble and hence lower effective OD. This effect, together with the intrinsic transverse expansion of OAM modes, is identified as the primary reason for the observed reduction in storage efficiency at higher mode orders. To further support this interpretation, we perform numerical simulations of storage efficiency as a function of mode index, taking into account spatial overlap with the atomic ensemble as shown in Fig.~\ref{figslife}(h). The results agree well with the experimental trends, confirming that the efficiency reduction is predominantly governed by spatial-mode-dependent OD rather than spin-wave mapping limitations.}

Crosstalk error in our setup originates from imperfect phase encoding and decoding by the SLMs. We measured the crosstalk using weak coherent probe pulses in the POV modes employed in the main text, with parameters $k_r=1.5$ and $\ell_i=-5,\dots,+5$. The phase pattern of SLM~1 (SLM~2) was set to $\ell_{in}$ ($\ell_{out}$). 
The photon counts were normalized to quantify the crosstalk error as
\begin{equation}
\label{eqACTE}
\begin{aligned}
CE_{i,j}=2N_{i,j}/(N_{i,i}+N_{j,j}),
\end{aligned}
\end{equation}
where $CE_{i,j}$ is the crosstalk error for $\ell_{in}=i,$ and $ \ell_{out}=j, (i,j=-5,\dots,+5)$, $N_{i,j}$ is the measured counts with $\ell_{in}=i, \ell_{out}=j$, and $N_{i,j}$ is the measured photon counts for the corresponding input and output modes, and $N_{i,i}$ ($N_{j,j}$) is the photon count when $\ell_{in}=\ell_{out}=i~(j)$.
The measured crosstalk is on average $0.4\%$, with a maximum value of $2.5\%$ observed for $\ell_{in}=+5, \ell_{out}=+4$, as shown in Fig.~\ref{figscro}. 
{We attribute this crosstalk mainly to the finite diffraction efficiency of SLMs and background noise during storage. The measured diffraction efficiency of the SLMs is $\eta_S=0.95$, leading to a residual unmodulated-photon contribution of $P_S=(1-\eta_S)^2/[\eta_S^2+(1-\eta_S)^2]\approx0.28\%$. This represents an additional noise contribution compared to previous results with qubit storage fidelity $F_s=99.6\%$~\cite{wang2019efficient}. In addition, we measured a scattered-photon background noise of approximately $0.3\%$ during the storage process by blocking the anti-Stokes photons and recording counts within the same time window. Regarding possible improvements, higher channel capacity and reduced crosstalk can be achieved by employing SLMs with higher diffraction efficiency and improved phase fidelity, thereby reducing unmodulated leakage, and optimizing the choice of encoding basis to avoid strongly overlapping neighbor modes.}

\begin{figure}[h]
\includegraphics[width=8.6cm]{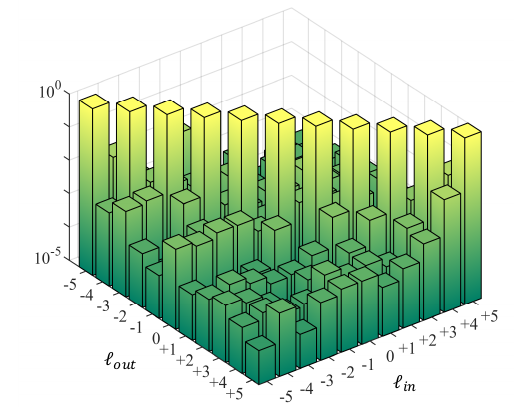}
\caption{{{Calibration of crosstalk error}. The matrix is constructed from POV modes by collecting photon counts for different input and output bases. The diagonal elements are normalized to unity, while the off-diagonal elements represent the crosstalk error rates between modes.}
}\label{figscro}
\end{figure} 

\section{III. Fidelity of qudits}

Here we provide the measurement details for determining the fidelity of a qudit state, following the method described in Refs.~\cite{pu2018experimental, PhysRevLett.124.240504}. For each spatial mode, the heralded anti-Stokes photon counts include contributions from accidental coincidences, which can be expressed as
\begin{equation}
\label{eqAScount}
\begin{aligned}
N^{(i)}_{AS}=\mathcal{h}^{(i)}N^{(i)}_{S}+N^{(i)}_{ac},
\end{aligned}
\end{equation}
where $N^{(i)}_{S}$, $N^{(i)}_{AS}$, and $N^{(i)}_{ac}$ are counts of Stokes photon, anti-Stokes photon, and accidental coincidence counts of the $i$-th spatial mode respectively. Thereby, the heralding rate of each POV mode can be extracted by $\mathcal{h}^{(i)}=(N^{(i)}_{AS}-N^{(i)}_{ac})/N^{(i)}_{S}$.
{In the experiment, $N_S$ corresponds to $N_1$ recorded by SPCM1. $N_{AS}$ denotes the total anti-Stokes photon counts, corresponding to the sum of $N_2$ and $N_3$ within the 800-ns temporal window. $N_{ac}$ represents the accidental coincidence counts measured by SPCM2 and SPCM3 using the same 800-ns window outside the anti-Stokes photon temporal profile.} 

The population of the qudit state includes contributions from both single- and double-photon excitations from the source. Therefore, the total excitation can be expressed as
\begin{equation}
\label{eqPopu}
\begin{aligned}
p_1+2p_2=\sum_{i}q^{(i)}/\mathcal{h}^{(i)}.
\end{aligned}
\end{equation}
The contribution of double excitation can be quantified by the anti-correlation, of which a simple relation with single and double excitation is written as 
\begin{equation}
\label{eqAntiCo}
\begin{aligned}
g_c^{2}(0)\approx\frac{2p_2}{p_1^2}.
\end{aligned}
\end{equation}
Combining Eq.~\ref{eqPopu} and Eq.~\ref{eqAntiCo} with the normalilzation $p_0+p_1+p_2=1$, we can obtain the population $p_0$, $p_1$ and $p_2$ from the measured $\mathcal{h}^{(i)}$, $q^{(i)}$ and $g_c^{(2)}(0)$. 

The state fidelity can be calculated by measured qudit population, $q_f$, which is bounded by
\begin{equation}
\label{eqFide}
\begin{aligned}
F\geq\frac{p_1q_f}{T(p_1+2p_2)}\left(\sum_i\sqrt{t'_i/11}\right)^2,
\end{aligned}
\end{equation}
where $T=\sum_i\mathcal{h}^{(i)}/11$ and $t'_i=\mathcal{h}^{(i)}/11T$.

\begin{figure}[h]
\includegraphics[width=8.6cm]{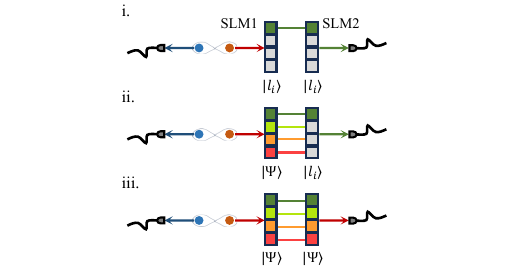}
\caption{{Fidelity measurement of qudits.} The lower bound of the qudit fidelity is determined through three steps. Heralded biphotons are generated in MOT~1, where detection of a Stokes photon (blue) projects the correlated anti-Stokes photon (red) into a single-photon state.
\textbf{i}, Measurement of heralding rates $\mathcal{h}^{(i)}$: SLM~1 encodes the anti-Stokes photon in the POV state $\ket{\ell_i}$, and SLM~2 decodes the same spatial mode $\ket{\ell_i}$.
\textbf{ii}, Measurement of qudit population distribution $q^{(i)}$: SLM~1 encodes the anti-Stokes photon in the 11-dimensional qudit state $\ket{\Psi}$, and SLM~2 projects onto the spatial modes $\ket{\ell_i}$ to measure the population in each mode.
\textbf{iii}, Measurement of the qudit population $q_f$: The full qudit phase pattern $\ket{\Psi}$ is applied to both SLMs to characterize the overall population fidelity. 
}\label{figswproc}
\end{figure} 

\section{IV. Channel capacity and Entanglement of formation} 
{In the main article, we consider the depolarizing noise $p_n$as the dominant source of storage infidelity, equally distributed among all modes, such that $p(y|x)=p_n/M,(y\ne x)$.
To compute the channel capacity, we assume equal input probabilities $p(x)=1/M$, and identical storage efficiencies for all modes $p(y|x)\equiv const,(y=x)$.}
Then the channel capacity can be written as, 
\begin{equation}
\label{eqCQM}
\begin{aligned}
C_{qm}=\log_2{M}+\left(1-\frac{M-1}{M}p_n\right)\log_2{\left(1-\frac{M-1}{M}p_n\right)}+(M-1)\frac{p_n}{M}\log_2{\frac{p_n}{M}}.
\end{aligned}
\end{equation}
From equation above, it is clear that any crosstalk error would reduce the channel capacity.

The performance of our quantum memory is quantified by storing and retriving photonic qudits, meanwhile, a practical quantum repeater channel is intended to distribute entanglement over long distances. Each of them stands for a point of view on quantum communication: either using a quantum channel to send qubits or using maximully entangled states for quantum teleportation. Nevertheless, the \textit{Quantum Reverse Shannon Theorem} states that a quantum channel can be simulated by shared entanglement with assisted classical communication~\cite{bennett2002entanglement}. 

In the following, we would like to check the impact of a quantum memory on channel capacity and entanglement. The channel capacity of a quantum memory can be calculated by {Eq.~2} in main article text.
To assess the effect of an imperfect multimode quantum memory on entangled qudit-qudit states, we evaluate the entanglement of formation, $E_F$. The qudit-qudit entangled state can be expressed as
\begin{equation}
\label{eqEQ}
\begin{aligned}
\ket{\Psi_M}=\frac{1}{\sqrt{M}}\sum_{i=1}^{M}\ket{\ell_i,\ell_i},
\end{aligned}
\end{equation}
thus the density matrix of an $M^2$-dimensional entangled state is $\rho_M=\ket{\Psi}\bra{\Psi}$. If we account for depolarizing noise $p_n$ arising from imperfect quantum memories, the retrieved entangled state can be expressed as 
\begin{equation}
\label{eqrhom}
\begin{aligned}
\rho'_M=(1-p_n)\rho_M+\frac{p_n}{M^2}I_M\otimes I_M.
\end{aligned}
\end{equation}
The entanglement of formation gives the minimal number of maximally entangled qubit-qubit states (ebits) that is required to get one copy of the desired entangled state through local operations and classical communication.
For a quantum memory with $M$ modes and $p_n$ depolarizing noise, a lower bound on the $E_F$ can be expressed as~\cite{PhysRevA.96.040303, PhysRevA.101.032312},
\begin{equation}
E_F\ge- \log_2(1-B^2/2),
\end{equation}
where $B$ is defined as
\begin{align}
B=\frac{2}{\sqrt{|C|}} \sum_{(\ell_j,\ell_k) \in C\atop j< k} &\Big( |\langle \ell_j,\ell_j|\rho'_M|\ell_k,\ell_k\rangle|-\sqrt{\langle \ell_j,\ell_k|\rho'_M|\ell_j,\ell_k\rangle\langle \ell_k,\ell_j|\rho'_M|\ell_k,\ell_j\rangle} \Big).
\end{align}
In the above equation, $j$ and $k$ denote the mode $\ell_j$ $\ell_k$ of the two entangled photonic qudits; $C$ is a set of mode pairs $(\ell_j,\,\ell_k)$ and $|C|$ denotes the number of pairs in the set.
Figure~\ref{figsCE} shows the calculated channel capacity and entanglement of formation, both of which indicate that a quantum memory with high storage fidelity is essential for enhancing the capacity of a quantum communication channel.

\begin{figure}[h]
\includegraphics[width=8.6cm]{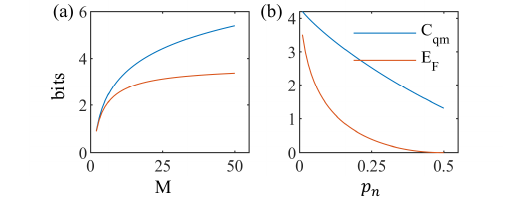}
\caption{{Channel capacity and entanglement of formation}. 
(a), Calculated $C_{qm}$ and $E_F$ with varing multimode number $M$ and fixed dipolarizing noise $p_n=2\%$.
{(b)}, Calculated $C_{qm}$ and $E_F$ with varing dipolarizing noise $p_n$ and fixed multimode number $M=20$
}\label{figsCE}
\end{figure}

\section{V. Comparation of quantum memories} 
{We use QIR to quantify the performance of our quantum memory as well as the others in this section. We suppose that only depolarizing noise induces storage infidelity, and most of quantum memories have tested their performance of storing photonic qubits with fidelity $F_s$, therefore, the depolarizing noise is estimated by the photonic qubit storage fidelity, which follows $F_s=1-p_n/2$. The channel capacity $C_{qm}$ can be estimated following Eq.~\ref{eqCQM}, and combining the distribution time $T_{tot}$ (Eq.~3 in main article) we can calcultate QIR by,
\begin{equation}
\label{eqRqm}
\begin{aligned}
 \mathcal{R}_{qm} = C_{qm}/T_{tot}.
\end{aligned}
\end{equation}}

Storage lifetime is as well a critical criterion for a quantum memory. For a realistic memory with finite lifetime $\tau$ (in unit of second), the average entanglement distribution time becomes~\cite{PhysRevLett.98.060502}, 
\begin{equation}
\label{eqTtau}
\begin{aligned}
T_{\tau}=\frac{T_{tot}-\frac{1+p_e}{p_ep_s}\frac{q^{\tau+1}}{1-p_e/2}}{1-\frac{q^{\tau+1}}{1-p_e/2}},
\end{aligned}
\end{equation}
where $p_s$ is probability of a successful entanglement swapping within a repeater node, $p_e$ is the probabilty to entangle adjacent memories, and $q=1-p_e$. After considering the storage time, the comprehensive performance could be modifies as, 
\begin{equation}
\label{eqRt}
\begin{aligned}
 \mathcal{R}_{\tau} = C_{qm}/T_{\tau}.
\end{aligned}
\end{equation}

In Table~\ref{tab1}, we benchmark the performance of our quantum memory, as well as other state-of-the-art multimode/multipartite quantum memories, using the criterion $\mathcal{R}_{qm}$ and $\mathcal{R}_{\tau}$. The following ideal assumptions are made in the comparison: 
(i) the total length of quantum repeater channel is 1000~km and the segement length is 250~km;
(ii) In all cases, storage efficiency refers to the intrinsic retrieval efficiency; 
(iii) multiplexing is counted only when each mode can be individually controlled or detected. Otherwise, modes are treated as degrees of freedom for a high-dimensional communication channel; 
(iv) the dominant error is assumed to arise from crosstalk between neighboring modes which is modeled as depolarizing noise;
(v) all modes are equally likely, i.e., $p(x)=1/M$ for $x=1,\dots,M$.

Under these conditions, our quantum memory outperforms others primarily due to its exceptionally high storage efficiency and fidelity, combined with intermediate multimode capacity. Its performance could be further enhanced by integrating additional multiplexing modes. Notably, the channel capacity increases slower with larger mode number when the storage fidelity is not sufficiently high~\cite{PhysRevLett.118.110501, PhysRevA.101.032312}. This highlights that, beyond merely increasing multimode capacity, high fidelity is also a decisive factor in achieving superior comprehensive performance.

\begin{table*}[]
\caption{{Performance of quantum memories}. {The comprehensive criteria  $\mathcal{R}_{qm}$ and $\mathcal{R}_{\tau}$ (in unit of minute) are used to estimate and compare different types of quantum memories. The performance of quantum memory are considered, including storage efficiency $\eta_s$, storage lifetime $\tau$, multiplexing mode number $N$, dimension mode number $M$, and storage fidelity $F_S$. The channel capacity $C_{qm}$ (in unit of bit) and distribution time $T_{tot}$/$T_{\tau}$ (without/with consideration of storage lifetime, in unit of minute) are estimated to get QIR.}
We assume ideal detection efficiency ($\eta_d=1$) and perfect entanglement swapping ($p_s=1$). 
For absorptive-type memories, a deterministic photon source is assumed with emission probability $p_e=0.7$~\cite{ding2025high}. 
For emissive-type memories, the emission probability is set to $p=0.1$ to suppress multiple excitations. 
The entanglement generation rate and retrieval efficiency are supposed the same $p=\eta_s=0.7$ in Ref.~\cite{hartung2024quantum}.
The fiber attenuation length is set as $L_{att}=22km$, corresponding to photon loss 0.2 dB/km.
}
\begin{ruledtabular}
\begin{tabular}{cccccccccccc}
Ref.&Photon source& $\eta_s$& $\tau$& N& M& $F_S$& $C_{qm}$& $T_{tot}$& $T_{\tau}$&$\mathcal{R}_{qm}$ &$\mathcal{R}_{\tau}$\\ \hline
This work&Single photons & $82.2\%$ & $28.0~\mu s$ & 1 & 11&$99.3\%$& 3.32& 0.93& 1.70&3.56& 1.99\\
\cite{wang2019efficient}&Single photons & $86.1\%$ &$10~\mu s$ & 1&2&$99.6\%$& 0.96& 0.69& 1.25&1.39& 0.78 \\
\cite{PhysRevLett.131.240801}&Weak pulses & $58\%$ &$4~\mu s$ & 1&25&$99.6\%$& 4.54& 4.54& 6.69&0.68& 0.37 \\
\cite{yang2025efficientmultiplex}&Weak pulses & $74.4\%$ &$5~\mu s$ & 4&4&$89.5\%$& 1.12& 0.43& 0.76&2.63& 1.48  \\
\cite{wei2024quantum} &Weak pulses& $2.8\%$ &$230~ns$ & 5&330&$99.9\%$& 8.33& 5.9e5& 1.8e4&1.41e-5& 7.57e-6  \\
\cite{teller2025solid} &Weak pulses& $5.7\%$ &$25~\mu s$ &10&250&$94\%$& 6.49& 1.6e4& 5.1e2&3.90e-4& 2.10e-4\\
\cite{PhysRevX.14.021018}&DLCZ & $25\%$ &$650~\mu s$ & 72&4&$95\%$& 1.50& 33.4& 62.0&4.48e-2& 2.42e-2  \\
\cite{hartung2024quantum}&Cavity-QED &$70\%$ &$1.1~ms$ & 6&2&$96.2\%$& 0.77& 0.40& 0.77&1.90& 1.07
\label{tab1}
\end{tabular}
\end{ruledtabular}
\end{table*}

\end{widetext}
\end{document}